\documentclass[nonstructabstract]{aa}
\usepackage{natbib}
\bibpunct{(}{)}{;}{a}{}{,}

\usepackage{graphicx}
\usepackage{textgreek}
\usepackage[caption=false]{subfig}
\usepackage{txfonts}
\usepackage{bm}
\usepackage[colorlinks=true, linkcolor=blue, citecolor=blue, urlcolor=blue]{hyperref}

\begin{document} 

    \titlerunning{SIMTERFERE}
    \authorrunning{J. R. Sauter et al.}

    \title{SIMTERFERE: An optical interferometry simulator for quantifying the coherent flux stability of VLTI/GRAVITY+}
    
    \subtitle{Reaching per mill stability: Application to exoplanet spectroscopy}

   \author{J. R. Sauter\inst{1,2}, A. von Stauffenberg\inst{1,2}, G. Bourdarot\inst{3}, W. Brandner\inst{1}, F. Eisenhauer\inst{3}, L. Kreidberg\inst{1}, L. Labadie\inst{4}, S. Scheithauer\inst{1}, D. Trevascus\inst{1,2}, R. van Boekel\inst{1}}

   \institute{Max-Planck-Institut für Astronomie, Königstuhl 17, 69117 Heidelberg, Germany\\
              \email{sauter@mpia.de}
            \and
            Fakultät für Physik und Astronomie, Universität Heidelberg, Im Neuenheimer Feld 226, 69120 Heidelberg, Germany
            \and
            Max-Planck-Institut für extraterrestrische Physik, Gießenbachstraße 1, 85748 Garching bei München, Germany
            \and
            I. Physikalisches Institut, Universität zu Köln, Zülpicher Str. 77, 50937 Köln, Germany
             }

   \date{}
 
  \abstract
   {The implementation of the GRAVITY+ Adaptive Optics (GPAO) system at VLTI enables unprecedented sensitivity and stability in optical interferometry. This allows high-precision characterization of directly imaged exoplanets at medium spectral resolution, providing a new pathway for studying planetary atmospheres.
   We aim to quantify and characterize the short- and long-term stability of GRAVITY+ through a consecutive seven-hour observation of the bright and stable star \textbeta\ Pictoris, providing a benchmark for future exoplanet observations.
   We developed \texttt{SIMTERFERE}, a data-driven simulation tool that reproduces GRAVITY+ on-star observations using ancillary instrument and telemetry data. By comparing the simulations with the measured coherent fluxes, we traced the origins of systematic flux variations and assessed their impact on exoplanet contrast measurements.
   We find that the $\sim 10\%$ variations are dominated by throughput changes driven by variable fiber coupling, which depends on wavefront stability, atmospheric dispersion, and residual fiber offsets. These variations appear as smooth continuum changes across wavelength and can be effectively mitigated using second-order polynomial corrections. After removing these instrumental effects, the remaining $\sim 1\%$ variations are almost purely of telluric origin, which we can reliably correct down to the photon-noise limit (0.1\% precision) using a contrast spectrum approach with linear airmass interpolation.
   The GRAVITY+ inferometric instrument is highly stable: low-order continuum and telluric variations can be corrected with high precision, making it uniquely capable of high-fidelity characterization of directly imaged exoplanets.}


   \maketitle


\section{Introduction} \label{section: introduction}
The GRAVITY instrument \citep{gravity_collaboration_first_2017} at the Very Large Telescope Interferometer (VLTI) enables adaptive optics (AO)-corrected, phase-referenced, near-infrared interferometric beam combination. During its first five years of operation, GRAVITY has revolutionized optical interferometry, enabling high-spatial-resolution studies of the Galactic Center, active galactic nuclei, exoplanets, and young stellar objects \citep{gravity_collaboration_detection_2018,gravity_collaboration_detection_2018-1,gravity_collaboration_spatially_2018,gravity_collaboration_gravity_2019}.

This success motivated a series of upgrades unified under the GRAVITY+ project \citep{gravity_collaboration_gravity_2022}, which aims to significantly improve sky coverage, sensitivity, and stability. A central component of this upgrade is the implementation of the new GRAVITY+ Adaptive Optics (GPAO) system \citep{gravity_collaboration_first_2025} on the four 8\,m unit telescopes (UTs). The GPAO system provides state-of-the-art extreme-AO wavefront correction using visible or infrared natural and laser guide stars. Initial tests demonstrate a factor-of-two improvement in the Strehl ratio.

Measuring the emission spectrum of an exoplanet in the presence of its much brighter host star remains extremely challenging because of the high contrast (typically 10\textsuperscript{-4} - 10\textsuperscript{-6}) that direct imaging methods must overcome. Detectability is closely tied to the achievable angular resolution, which dictates the size of the unresolved stellar point spread function (PSF). For a diffraction-limited telescope of aperture size $D$, the angular resolution scales as $\lambda/D$, while for an interferometer it scales as $\lambda/2B$, where $B$ is the longest baseline, providing an order-of-magnitude improvement in resolving power \citep{eisenhauer_advances_2023}. However, GRAVITY has not yet fully exploited its 130-m baselines because sensitivity remained dominated by stellar photon and speckle noise coupled into the single-mode fiber. The improved performance of GPAO substantially reduces these noise sources and enhances the sensitivity to exoplanet emission by approximately an order of magnitude \citep{gravity_collaboration_first_2025}.

These advancements enable precise and high-cadence spectral characterization of exoplanets using GRAVITY’s high-resolution grating ($R \approx 4000$). The K-band (1.97\,µm – 2.40\,µm) offers a unique spectral window to probe the chemical and cloud structure of young gas giants, placing the first isotopologue detections and variability measurements of directly imaged exoplanets within reach. Such detections offer transformative insights into exoplanet formation histories and atmospheric dynamics \citep{gandhi_jwst_2023,mulder_eso_2025,biller_jwst_2024,chen_jwst_2025}.

Both isotopologue and atmospheric-variability measurements require multi-hour integrations to achieve high signal-to-noise ratios (S/Ns) and to track hour-scale changes in atmospheric structure \citep{ravet_multi-modal_2025}. While GRAVITY+ demonstrates the sensitivity required for these observations, the spectrometric stability over multi-hour timescales remains uncharacterized.

The recent upgrade of GRAVITY to GRAVITY+ opens new and promising opportunities for ground-based exoplanet characterization. However, as the upgraded instrument was only recently deployed, its actual performance and limitations remain largely unexplored.

Recent analyses of exoplanet spectra obtained with GRAVITY reveal flux variations at the level of approximately 10\% between multi-epoch observations. These variations are attributed to instrumental effects that are not yet fully understood \citep{kammerer_exogravity_2025}, which limits the reliability of spectroscopic measurements and their physical interpretation.

The primary goal of this work is to characterize the spectrometric flux stability of GRAVITY+ by systematically investigating the extent and properties of instrumental flux variations. In addition, we aim to identify the underlying causes of these variations and to develop and validate mitigation strategies within the \texttt{exogravity} data-reduction framework.

We developed a data-driven simulation tool that reproduces GRAVITY+ on-star observations using ancillary instrument and telemetry data. This approach isolates individual instrumental processes, enabling detailed analysis and interpretation of systematic effects present in the data. The resulting framework provides a flexible diagnostic tool that is publicly available to the community for targeted performance studies.

A deeper understanding of the spectrometric stability and performance of GRAVITY+ is essential to improve current observations and guide future instrument development and observing strategies.

The \texttt{exogravity}\footnote{\url{https://gitlab.obspm.fr/mnowak/exogravity}} pipeline \citep{gravity_collaboration_first_2019,nowak_direct_2020,nowak_catalogue_2024} utilizes the high spatial resolution of GRAVITY to detect and characterize directly imaged exoplanets. Following the notations of \cite{balmer_vltigravity_2025}, the measured interferometric observable is the coherent flux $\Gamma$ of an object at baseline $\bm{\mathsf{b}}$, time $t$, and wavelength $\lambda$ is
\begin{align}\label{equation: coherent flux}
   \Gamma(\bm{\mathsf{b}},t,\lambda) = F(t,\lambda) \cdot \mathcal{V}(\bm{\mathsf{b}},\lambda) \cdot e^{\mathrm{i}\frac{2\pi}{\lambda}[\alpha\cdot U(\bm{\mathsf{b}}) + \delta \cdot V(\bm{\mathsf{b}})]},
\end{align}
where $F \in \mathbb{R}$ is the spectral flux, $\mathcal{V} \in \mathbb{C}$ is the complex visibility, ($\alpha, \delta) \in \mathbb{R}^2$ is the angular sky position, and $(U,V) \in \mathbb{R}^2$ are the projected baseline coordinates in the uv-plane.

The flux contrast $C \in (0,1)$ between the planet $p$ and the star $s$ is given by
\begin{align}
   C(t,\lambda) &= \frac{F_{p}(t,\lambda)}{F_{s}(t,\lambda)} \nonumber\\
   &= \frac{\Gamma_{p}(\bm{\mathsf{b}},t,\lambda)}{\Gamma_{s}(\bm{\mathsf{b}},t,\lambda)}\frac{\mathcal{V}_{s}(\bm{\mathsf{b}},\lambda)}{\mathcal{V}_{p}(\bm{\mathsf{b}},\lambda)}e^{\mathrm{i}\frac{2\pi}{\lambda}[\Delta\alpha\cdot U(\bm{\mathsf{b}}) + \Delta\delta \cdot V(\bm{\mathsf{b}})]}.
\end{align}
The observed stellar coherent flux $\Gamma_{os} \in \mathbb{C}$ is modified by atmospheric and instrumental transmissivity $G \in (0,1)$:
\begin{align}\label{equation: observed coherent flux}
   \Gamma_{os}(\bm{\mathsf{b}},t,\lambda) = G(\bm{\mathsf{b}},t,\lambda) \cdot \Gamma_{s}(\bm{\mathsf{b}},t,\lambda).
\end{align}
The observed planetary coherent flux $\Gamma_{op}$ is additionally influenced by residual stellar leakage from the stellar PSF at the science channel (SC) fiber position. This leakage can be modeled as a low-order wavelength polynomial $Q_{\lambda} \in \mathbb{R}$ applied to the stellar coherent flux:
\begin{align} \label{equ: on planet}
   \Gamma_{op}(\bm{\mathsf{b}},t,\lambda) &= G(\bm{\mathsf{b}},t,\lambda) \cdot \Gamma_{p}(\bm{\mathsf{b}},t,\lambda) + Q(\bm{\mathsf{b}},t,\lambda)\cdot\Gamma_{os}(\bm{\mathsf{b}},t,\lambda) \nonumber\\
   &= \Gamma_{os}(\bm{\mathsf{b}},t,\lambda)\,\cdot \nonumber\\
   & \left[C(t,\lambda)\frac{\mathcal{V}_{p}(\bm{\mathsf{b}},\lambda)}{\mathcal{V}_{s}(\bm{\mathsf{b}},\lambda)}
   e^{-\mathrm{i}\frac{2\pi}{\lambda}[\Delta\alpha\cdot U(\bm{\mathsf{b}}) + \Delta\delta \cdot V(\bm{\mathsf{b}})]}
   + Q(\bm{\mathsf{b}},t,\lambda) \right].
\end{align} 
The \texttt{exogravity} pipeline determines the best-fitting position and contrast spectrum through chi\textsuperscript{2}-minimization:
\begin{align}
   \chi^{2}(\Delta\alpha,\Delta\delta,Q,C) = \left[\bm{\hat{\mathsf{\Gamma}}}_{op}-\bm{\mathsf{\Gamma}}_{op}\right]^{T} \times \bm{\mathsf{W}}^{-1} \times \left[\hat{\bm{\mathsf{\Gamma}}}_{op}-\bm{\mathsf{\Gamma}}_{op}\right],
\end{align}
where $\bm{\mathsf{\Gamma}}_{op}$ is the coherent flux model in matrix form, $\hat{\bm{\mathsf{\Gamma}}}_{op}$ are the measured data, and $\bm{\mathsf{W}}$ is the respective covariance matrix. In this work, we denote matrix multiplications as $\times$ and transpose as $^{T}$.

In practice, the star and the planet cannot be observed simultaneously in the SC. The SC fiber is therefore periodically switched between the stellar and planetary positions. During on-planet observations, the on-star coherent flux is estimated as the average of the on-star measurements taken immediately before and after:
\begin{align}\label{equation: on-star prediction}
   \Gamma_{os}(\bm{\mathsf{b}},t,\lambda) &\approx \frac{\hat{\Gamma}_{os}(\bm{\mathsf{b}},t+\Delta t_2,\lambda)+\hat{\Gamma}_{os}(\bm{\mathsf{b}},t-\Delta t_1,\lambda)}{2} \nonumber\\
   &\approx \frac{G(\bm{\mathsf{b}},t+\Delta t_2,\lambda)+G(\bm{\mathsf{b}},t-\Delta t_1,\lambda)}{2}\Gamma_{s}(\bm{\mathsf{b}},\lambda). 
\end{align}
The modeled on-planet coherent flux in Eq. (\ref{equ: on planet}) is therefore only valid for the minimization if both the transmissivity and stellar flux vary slowly in time.

Hence, time-variable transmissivity directly biases the derived contrast spectrum, making it a strong candidate for the observed instrumental flux variation. This highlights the need to thoroughly characterize the transmissivity of GRAVITY and develop strategies to mitigate or correct its temporal variations.

\textbeta\ Pictoris is a young (23 $\pm$ 3 Myr), nearby (19.44 $\pm$ 0.05 pc), and therefore bright ($K = 3.48\,\mathrm{mag}$) intermediate-mass (1.75 M$_\odot$) A6V star that hosts a prominent edge-on debris disk and two confirmed gas-giant planets \citep{crifo__1997,ducati_vizier_2002,van_leeuwen_hipparcos_2007,mamajek_age_2014,nielsen_gemini_2020,nowak_direct_2020}. Although \textbeta\ Pictoris is a known \textdelta-Scuti variable star \citep{koen_pulsations_2003}, its variability is small $\lesssim 1\%$ in the visible \citep{koen_pulsations_2003} and $\ll 1\%$ in the K band.

A multi-hour observation of the \textbeta\ Pictoris system is therefore ideal for tracking the temporal behavior of atmospheric and instrumental transmissivity. \textbeta\ Pictoris' near-constant K-band brightness enables high-S/N measurements while ensuring that temporal variations in the coherent flux primarily reflect instrumental or atmospheric effects rather than intrinsic stellar variability.

\begin{figure*}[!t]
  \sidecaption
  \includegraphics[width=12cm]{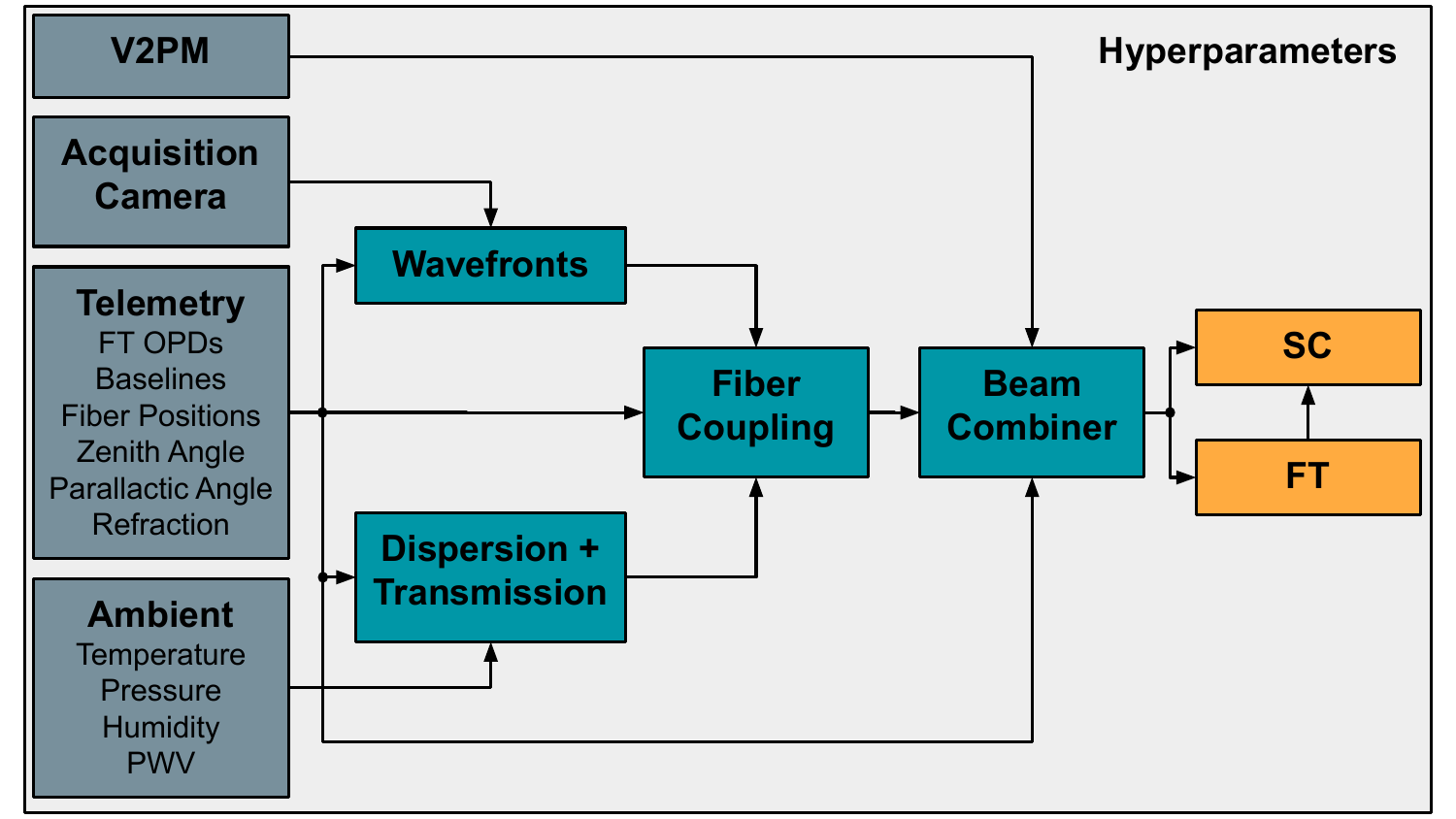}
  \caption{Flow chart of the \texttt{SIMTERFERE} simulation framework. Simulated coherent fluxes (orange) are generated by injecting atmospheric and instrumental effects (blue), derived from the ancillary telemetry of the GRAVITY observations (dark gray). The overall configuration of the simulation and the subcomponents is controlled through user-defined hyperparameters (light gray).}
  \label{figure:flow-chart}
\end{figure*}

The paper is organized as follows. Section \ref{section: data} summarizes our GRAVITY observations of the \textbeta\ Pictoris system and the corresponding data reduction. Section \ref{section: simterfere} introduces \texttt{SIMTERFERE}, the simulation framework used for our analysis and characterization of the data. We present the results in Sect. \ref{section: results}, and we discuss the limitations of the simulation tool in Sect. \ref{section: limitations}. Section \ref{section: discussion and conclusion} provides an overall discussion and concludes the paper.  

\section{Data} \label{section: data}
We observed the \textbeta\ Pictoris system on 20 December 2024 for a total of seven consecutive hours using GRAVITY in the Guaranteed Time Observations (GTO) program 114.27JS (PI: L. Kreidberg), covering airmasses from 1.1 to 1.5. We carried out the observations with the four UTs in high-spectral-resolution mode ($R \approx 4000$) and dual-field on-axis mode configuration. In this setup, the fringe tracker (FT) fiber remains locked on the star for fringe tracking and phase referencing, while the SC fiber alternates between the predicted planet position, the host star, and dedicated sky locations. Table \ref{table: exposure times} summarizes the exposure settings for the SC fiber pointings.

\begin{table}[h]
\caption{Exposure settings for the \textbeta\ Pictoris observations.}           
\label{table: exposure times}     
\centering                         
\begin{tabular}{c c c c c}        
\hline\hline                 
Target & \#Exp. & \#Int. & DIT [s] & Total [min]\\     
\hline                        
   $\beta$ Pic & 18 & 16 & 3 & 14.4\\     
   $\beta$ Pic b & 38 & 4 & 100 & 253.3\\
   sky a & 9 & 16 & 3 & 7.2 \\
   sky b & 10 & 4 & 100 & 66.7 \\
\hline        
\end{tabular}
\tablefoot{The different columns comprise the SC fiber pointings, total number of exposures, integrations per exposure, detector integration time (DIT), and combined total exposure times.}
\end{table}

As summarized in Table \ref{table: observation conditions}, the atmospheric conditions were generally good throughout the night. However, ground wind speeds were exceptionally low (<\,3\,m/s), causing the low-wind effect (LWE; \citealt{milli_low_2018,pourre_low-wind-effect_2022}), which degrades the quality and stability of the PSF. Sect. \ref{subsection: lwe} discusses the impact of the LWE on our data.

\begin{table}[h]
\caption{Atmospheric conditions during observations of \textbeta\ Pictoris at the Paranal astronomical site.}             
\label{table: observation conditions}      
\centering               
\begin{tabular}{ccc}  
\hline\hline 
Quantity & Value & Unit \\
\hline 
Seeing ($\lambda = 500\,\mathrm{nm}$) & $0.69^{+0.09}_{-0.08}$ & ["] \\  
Coherence time & $7.1^{+1.4}_{-1.3}$ & [ms] \\
Wind Speed & $1.0^{+0.7}_{-0.3}$ & [m/s] \\
Precipitable Water Vapor & $1.56^{+0.04}_{-0.07}$ & [mm] \\
\hline                                  
\end{tabular}
\tablefoot{Values are temporal medians with 16–84th percentile range uncertainties.}
\end{table}

We reduced the raw data using the ESO GRAVITY Public Release 1.9.4. pipeline and processed it up to the \texttt{astroreduced} intermediate data products. At this stage, individual integrations are kept separate rather than averaged, which is essential for our analysis of short-timescale stability. For on-star observations, the pipeline reports a median absolute coherent flux S/N of $180^{+60}_{-40}$ per spectral bin, integration, and baseline, with uncertainties corresponding to the 16–84th percentile range.

\section{\texttt{SIMTERFERE}} \label{section: simterfere}
The \texttt{SIMTERFERE}\footnote{\url{https://github.com/Manaffel/SIMTERFERE}} tool is a publicly available Python-based data-driven interferometric simulator designed to reproduce VLTI/GRAVITY on-star observations using their associated ancillary data. The GRAVITY instrument is highly complex, with numerous astronomical, atmospheric, and instrumental processes acting simultaneously. Consequently, features observed in the coherent flux or visibility are often the product of multiple coupled effects, making it challenging to attribute observed systematics to individual subsystems. The \texttt{SIMTERFERE} package addresses this challenge by providing controlled access to the various contributors to the measurement process, allowing targeted diagnostic tests.

Using GRAVITY's ancillary data, we can inject the behavior of key subsystems, such as AO \citep{arsenault_macao-vlti_2003,scheithauer_ciao_2016,gravity_collaboration_gravity_2022,gravity_collaboration_first_2025}, delay lines \citep{derie_vlti_2000,pfuhl_gravity_2012,drescher_gravity_2022}, and fringe tracker \citep{lacour_gravity_2019,nowak_upgrading_2024}, directly into the simulation without requiring detailed modeling. This approach allows \texttt{SIMTERFERE} to include most instrumental effects while remaining computationally efficient, modular, and easily extendable. 

Fig. \ref{figure:flow-chart} shows a schematic overview of the simulation workflow. The simulation mimics the propagation of stellar light through Earth’s atmosphere, telescopes, and the GRAVITY instrument, following the most prominent sequence of physical and instrumental effects that affect real observations.

Starting from a user-defined input spectrum, the simulation applies atmospheric transmission to account for wavelength-dependent absorption and throughput losses. The light is then injected into the single-mode fibers of the four telescopes, where the simulation determines the coupling efficiency from measured wavefront stability and chromatic dispersion effects. The four beams are subsequently combined into the six interferometric baselines, and the recorded residual piston errors are applied to reproduce the temporal phase perturbations observed during real observations.

In a final step, the simulation interferes with the light using a linear model of the GRAVITY beam combiner, which includes the overall instrumental throughput. This procedure is performed independently for the FT and SC channels, taking into account their different integration times and spectral resolutions. The resulting simulated observables can then be directly compared to actual GRAVITY+ data, allowing instrumental effects to be isolated and studied in a controlled manner.

Each component of the simulation workflow is described in more detail in the following sections. For a comprehensive description of GRAVITY and its subsystems, we refer the reader to the instrument papers cited above.

\subsection{Ancillary data} \label{subsection: ancillary data}
The GRAVITY ancillary data provides real-time measurements of atmospheric properties and instrumental performance, recorded as telemetry and ambient parameters.  Using these data streams as direct inputs, \texttt{SIMTERFERE} reproduces the dominant effects seen in the observations without requiring full end-to-end simulations of each subsystem.

\subsubsection{Visibility-to-pixel-matrix (V2PM)} \label{subsubsection: v2pm}
The visibility-to-pixel-matrix $\bm{\mathsf{V2PM}}(\lambda) \in \mathbb{C}^{6\times4\times10}$ \citep{tatulli_interferometric_2007} provides a linear description of the GRAVITY beam combiner at wavelength $\lambda$. The detected signal $\bm{\mathsf{P}}(t,\lambda) \in \mathbb{R}^{6\times4}_+$ at time $t$, corresponding to the coherent fluxes $\bm{\Gamma}(t,\lambda) \in \mathbb{C}^{6}$ of the six baselines and the fluxes $\bm{f}(t,\lambda) \in \mathbb{R}^{4}_+$ of the four telescopes, is given by
\begin{align}\label{equation: v2pm coherent flux}
   \bm{\mathsf{P}} = \bm{\mathsf{V2PM}}\times\left[\bm{\mathsf{f}},\bm{\mathsf{\Gamma}}\right].
\end{align}
For simplicity, we omit time and wavelength dependence in the following equations. The GRAVITY instrument employs the ABCD-modulation scheme \citep{lacour_characterization_2008}, in which the light from each baseline is split into four beams with different phase shifts (i.e., $A: 0$, $B: \pi/2$, $C: \pi$, and $D: 3\pi/2$, for an ideal beam combiner). The measured $6\times4=24$ signals result from the combination of the six baselines with these four ABCD outputs.

During calibration, the instrument uses an internal light source. For each baseline combination $i = 1, ..., 6$, only the respective two shutters are opened, and the 24 signals are measured while applying a phase modulation $\bm{\mathsf{\phi}}(t) \in (0,2\pi)^{6}$ through the delay lines. The calibration signals $\hat{\bm{\mathsf{P}}}(t,\lambda) \in \mathbb{R}_{+}^{6\times6\times4}$ are given by 
\begin{align} \label{Equation: V2PM Calibration}
    \hat{\mathsf{P}}_{i,j,k} 
    &= \mathsf{F}_{i,\mathsf{\alpha_j},k} + \mathsf{F}_{i,\mathsf{\beta}_j,k} + \notag \\
        &\quad 2\mathsf{\gamma}_{i,j,k}\sqrt{\mathsf{F}_{i,\mathsf{\alpha}_j,k}\mathsf{F}_{i,\mathsf{\beta}_j,k}}\cos\left(\phi_{i}+\Delta\mathsf{\Phi}_{i,j,k}\right),
\end{align}
where $\bm{\mathsf{\gamma}}(\lambda) \in (0,1)^{6\times6\times4}$ is the coherence, $\bm{\mathsf{\alpha}} = [1,1,1,2,2,3],\, \bm{\mathsf{\beta}} = [2,3,4,3,4,4]$ are the two interfering telescope indices for the six baselines $j = 1,...,6$, and $k = 1,...,4$ labels the four ABCD beams.

We obtain the V2PM by fitting the measured calibration signals to determine the fluxes $\bm{\mathsf{F}}(\lambda) \in \mathbb{R}_{+}^{6\times4\times4}$, phase shifts $\Delta\bm{\mathsf{\Phi}}(\lambda) \in (0,2\pi)^{6\times6\times4}$, and coherence $\bm{\mathsf{\gamma}}(\lambda) \in (0,1)^{6\times6\times4}$. We then define the V2PM as
\begin{align} \label{equation: v2pm}
    \mathsf{V2PM}_{i,k,l} = 
    \begin{cases}
      \mathsf{F}_{i,l,k} & l = 1,...,4 \\
    2\mathsf{\gamma}_{i,l-4,k}\sqrt{\mathsf{F}_{i,\mathsf{\alpha}_{l-4},k}\mathsf{F}_{i,\mathsf{\beta}_{l-4},k}}e^{\mathrm{i}\Delta\mathsf{\Phi}_{i,l-4,k}} & l = 5,...,10,
    \end{cases}
\end{align}
and finally normalized to ensure unit throughput response:
\begin{align}
    \mathsf{V2PM}^{*}_{i,k,l} = 
    \begin{cases}
      \frac{24\cdot\mathsf{V2PM}_{i,k,l}}{\sum_{i,k=1}^{6,4}\mathsf{V2PM}_{i,k,l}} & l = 1,...,4 \\
      \frac{24\cdot\mathsf{V2PM}_{i,k,l}}{\sqrt{\sum_{i,k=1}^{6,4}\mathsf{V2PM}_{i,k,\alpha_{l-4}}\sum_{i,k=1}^{6,4}\mathrm{V2PM}_{i,k,\beta_{l-4}}}} & l = 5,...,10.
    \end{cases}
\end{align}
During science observations, the simulation computes the pixel-to-visibility-matrix $\bm{\mathsf{P2VM}}(\lambda) \in \mathbb{C}^{10\times6\times4}$ as the pseudo-inverse of V2PM\textsuperscript{*} and applies to the measured signals to recover the coherent fluxes:
\begin{align}
   \left[\bm{\mathsf{f}},\bm{\mathsf{\Gamma}}\right] = \bm{\mathsf{P2VM}}\times\bm{\mathsf{P}}.
\end{align}
In \texttt{SIMTERFERE}, individual V2PM matrices can be defined separately for the SC and FT. We then apply the P2VM following the procedure described above.

\subsubsection{Acquisition camera} \label{subsubsection: acquisition camera}
The GRAVITY acquisition camera \citep{anugu_design_2014,anugu_gravity_2016} performs beam alignment, active pupil and field stabilization, defocus correction, and engineering tasks. It comprises a pupil tracker, a field tracker, a pupil imager, and an aberration tracker, all recorded simultaneously on a single detector with a cadence of 0.7\,s. 

A dichroic in each telescope beam separates the H-band flux, directing it to the acquisition camera. The field images have a pixel scale of $p_{AC} = 17.8\,mas$. 

In \texttt{SIMTERFERE}, the four PSF images are extracted using fixed windows of size $75 \times 75\,\mathrm{pixel}$.

\subsubsection{Telemetry and ambient parameters}
The GRAVITY instrument software subsystem (ISS) computes relevant telemetric data in real time, including the zenith angle (for atmospheric dispersion and transmission), fiber offsets (for fiber coupling), baselines and fringe-tracker optical path differences (OPDs; for the beam combiner), and the parallactic angle (for wavefront sensing). It records these data in the FITS headers of the observation files.

The METEO and LHATPRO \citep{querel_all-sky_2014} systems at the VLT measure ambient parameters in real time. The simulations use ambient temperature, pressure, relative humidity, and perceptible water vapor (PWV) to compute atmospheric dispersion and transmission effects. 

\subsection{Wavefronts} \label{subsection: wavefronts}
The extracted acquisition camera field images of the star serve as underlying PSF references for the applied focal-plane wavefront sensing. For each image, we subtracted the temporally closest median sky observation to remove bias, dark current, and background. The fitting routine limits the maximum count to 40,000\,ADU to mitigate saturation effects. 

We modeled the VLT aperture using \texttt{POPPY}\footnote{\url{https://github.com/spacetelescope/poppy}} \citep{perrin_poppy_2016} for wavefront reconstruction. Following \cite{por_high_2018}, we modeled the primary mirror as a \texttt{CircularAperture} with radius 4.1\,m, and the secondary mirror as an \texttt{AsymmetricSecondaryObscuration} with a radius of 0.558\,m for UT1-3 and 0.6495\,m for UT4. We placed four supporting spiders, each 0.04\,m wide, around the mirror at angles of 39.5°, 140.5°, 219.5°, and 320.5°. We extracted the aperture using \texttt{CompoundAnalyticOptic.get\_transmission} at $\lambda_H = 1.654\,\mu m$ and placed the 33-pixel aperture model in the center of a $75 \times 75\,\mathrm{pixel}$ zero-array to match the acquisition camera pixel scale at Nyquist sampling ($17.8\,\mathrm{mas} \cdot 0.5D/\lambda \approx 75/33$).  

We implemented the focal plane wavefront sensing following \cite{feng_exoplanet_2025} in \texttt{PyTorch} \citep{paszke_pytorch_2019}. The simulation rotates the spiders in the aperture $\bm{\mathsf{A}}_H(t) \in (0,1)^{75\times75}$ according to the parallactic angle. It fits the phase $\bm{\mathsf{\Phi}}_H(t) \in (0,2\pi)^{75\times75}$, amplitude $\mathrm{amp}(t) \in \mathbb{R}_+$, and uniform background $\mathrm{bg}(t) \in \mathbb{R}_+$ via $\chi^2$-minimization, assuming photon noise as the main error source in our loss function $\mathcal{L}(\mathrm{amp},\mathrm{bg},\bm{\mathsf{\Phi}}_H) \in \mathbb{R}_+$:

\begin{align}
    \mathcal{L} = \sum_{x=-37}^{37}\sum_{y=-37}^{37} \frac{\left(\mathrm{amp} \cdot \mathsf{PSF}_{x,y} + \mathrm{bg} - \widehat{\mathsf{PSF}}_{x,y}\right)^2}{\left|\widehat{\mathsf{PSF}}_{x,y}\right|}, 
\end{align}
where $\bm{\mathsf{PSF}}(t) = \left|\bm{\mathcal{F}}\left\{\bm{\mathsf{A}}_H \cdot e^{\mathrm{i}\bm{\mathsf{\Phi}}_H}\right\}\right|^2 \in \mathbb{R}_{+}^{75\times75}$ is the PSF model, with $\bm{\mathcal{F}}$ denoting the discrete 2D Fourier transform. The simulation saturates the modeled PSF at 40,000\,ADU to match the detector properties.

We initialized the minimization by first fitting a third-order 2D polynomial wavefront along with amplitude and background. We set the polynomial to zero, the amplitude to one, and the background to zero as an initial guess. The initial fit uses \texttt{torch.optim.AdamW} for 3000 iterations with a learning rate of 10\textsuperscript{-3}. The solution is then refined through a full phase minimization with 20,000 iterations, using the previous fit as the initial guess. For subsequent frames within an exposure, the simulation uses the solution from the previous frame as an initial guess to prevent phase jumps.

To extrapolate from H to K band, the simulation scales the K-band phase $\bm{\mathsf{\Phi}}_K(t) \in (0,2\pi)^{99\times99}$ and aperture $\bm{\mathsf{A}}_K(t) \in (0,1)^{99\times99}$ by $\lambda_K/\lambda_H \approx 99/75$ by zero-padding at the edges. 

Because the acquisition camera, fringe tracker, and science cadences are not synchronized, the simulation linearly interpolated the unwrapped phases for intermediate time steps.

Notably, the focal-plane wavefront reconstruction has sign ambiguity, meaning that the solution is not unique. Nevertheless, the obtained solutions reproduce statistical properties and PSF shapes, accurately modeling the effective throughput losses in the fiber coupler.

\subsection{Dispersion} \label{subsection: refraction}
The simulation calculates atmospheric dispersion using an extension of the Cassini approximation \citep[e.g.][]{young_sunset_2004}, modeling a two-layer atmosphere with a moist layer at the observatory level and a dry layer above. We set the height of the moist layer to 2\,km and adopted the refractive indices of moist and dry air from \cite{mathar_refractive_2007}. We set the ambient conditions of the observatory to the mean values during the observations: a temperature of 14.41°C and a pressure of 744.7\,hPa.

We precomputed a grid of diffraction curves for PWV values from 0.5\,mm to 2.5\,mm in 16 increments, and zenith angles from 39.7° to 54.0° in eight increments. We performed cubic interpolation of this regular grid using \texttt{RegularGridInterpolator} from \texttt{SciPy} \citep{virtanen_scipy_2020} to retrieve the relative dispersion $(\mathrm{disp}_x,\mathrm{disp}_y)(t,\lambda) \in \mathbb{R}^2$ on the detector, where we adopted the detector xy-directions from the recorded telemetry refraction values. 

\subsection{Transmission} \label{subsection: transmission}
We computed the atmospheric transmission $G_{\mathrm{telluric}}(t,\lambda) \in (0,1)$ using \texttt{LBLRTM}\footnote{\url{https://github.com/AER-RC/LBLRTM}} via the Python wrapper \texttt{TelFit}\footnote{\url{https://github.com/kgullikson88/Telluric-Fitter}} \citep{clough_line-by-line_1992,clough_atmospheric_2005,gullikson_correcting_2014}. Following Eq. (\ref{equation: coherent flux}) and (\ref{equation: observed coherent flux}), the transmitted flux $f_{\mathrm{telluric}}(t,\lambda) \in \mathbb{R}_+$ is then given as
\begin{align}
    f_{\mathrm{telluric}} = G_{\mathrm{telluric}}f_{s},
\end{align}
where $f_{s}(\lambda) \in \mathbb{R}_+$ denotes the input stellar spectrum.

We initialized the transmission using the Paranal observatory altitude (2635\,m) and latitude (24°37'39"\,S). We used the recorded ambient pressure, temperature, relative humidity, and zenith angle as input parameters. The relative abundance of other molecular species (e.g. CO\textsubscript{2}, CH\textsubscript{4}, and CO) can either be set manually or retrieved using the fitting options of \texttt{LBLRTM}/\texttt{TelFit}.

\begin{figure*}[!t]
    \sidecaption
    \includegraphics[width=12cm]{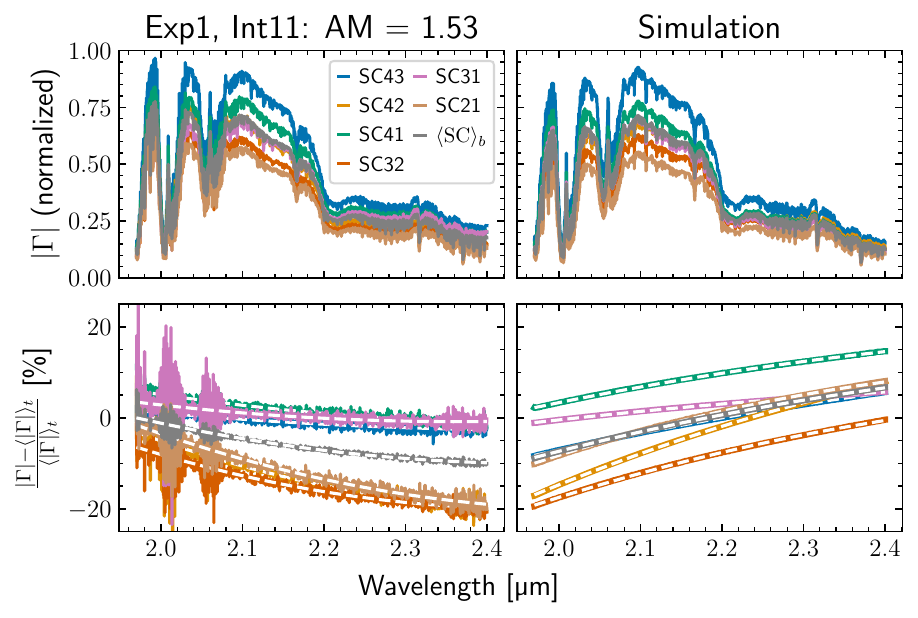}
    \caption{\textit{Top:} Example frame (exposure 1/18, integration 11/16) showing measured (left) and simulated (right) absolute SC coherent flux per wavelength bin for the six baselines (43, ..., 12) and their baseline averages $\langle \mathrm{SC} \rangle_b$ (different colors), taken at high airmass. Absolute coherent fluxes are normalized to the maximum flux in the exposure. \textit{Bottom:} Relative absolute coherent flux differences with respect to the time-averaged value for each baseline. Second order polynomial fits are indicated by dashed white lines. The systematic differences between the measured and simulated coherent fluxes are attributed to unknown fiber offsets, which are discussed in more detail in Sect. \ref{subsection: fiber offsets}. The associated movie is available online.}
    \label{figure: visibility_example}
\end{figure*}

\subsection{Fiber coupling} \label{subsection: fiber coupling}
We simulated the fiber as an ideal single-mode fiber using a Gaussian profile, with the fiber FWHM given by \citep{perrin_single-mode_2019}\begin{align}
    \mathrm{FWHM}_{K} = \frac{\lambda_K}{D}.
\end{align}
We computed the fiber profile $\bm{\mathsf{fp}}(t,\lambda) \in \mathbb{C}^{99\times99}$ in pupil space as
\begin{align}
    \bm{\mathsf{fp}} = e^{-2\pi^2\sigma^2_{\mathrm{fiber}}\left(\bm{\mathsf{U}}^2+\bm{\mathsf{V}}^2\right) +2\pi \mathrm{i}\left(x_{\mathrm{fiber}}\bm{\mathsf{U}}+y_{\mathrm{fiber}}\bm{\mathsf{V}}\right)},
\end{align}
where the fiber width $\sigma_{\mathrm{fiber}}$ scales continuously with wavelength (instead of the discretized aperture size):
\begin{align}
    \sigma_{\mathrm{fiber}} = \frac{\mathrm{FWHM}_{K}}{2\sqrt{2\ln{2}}}\frac{\lambda_K}{\lambda}. 
\end{align}
Here, $\bm{\mathsf{U}},\bm{\mathsf{V}} \in \mathbb{R}^{99x99}$ are the detector coordinates in pupil space and $x_{\mathrm{fiber}}/y_{\mathrm{fiber}}(t) \in \mathbb{R}$ are the recorded fiber positions.

We calculated flux coupling $\eta(t,\lambda) \in (0,1)$  as 
\begin{align}
    \eta = \left|\sum_{x=-49}^{49}\sum_{y=-49}^{49}\bm{\mathsf{fp}}^\dagger_{x,y}\mathsf{A}_{K,x,y}e^{\mathrm{i}\left(\frac{\lambda_H}{\lambda}\mathsf{\Phi}_{K,x,y}-\frac{2\pi}{99p_{AC}}[x \cdot \mathrm{disp}_x + y \cdot\mathrm{disp}_y]\right)}\right|.
\end{align}

\subsection{Beam combiner} \label{subsection: beam combiner}
We computed the detected signals from the four telescope coupled fluxes $\bm{\mathsf{\eta}}(t,\lambda) \in (0,1)^4$ using the beam combination described by the V2PM components. Applying Eq. (\ref{equation: v2pm coherent flux}) and (\ref{equation: v2pm}), the output signals are given as
\begin{align} \label{equation: output signals}
    \mathsf{P}_{j,k} &= \sum_{i=1}^6 \left(1 - \mathsf{\xi}_{i,j,k}\right) \cdot \left(\mathsf{I}_{i,\mathsf{\alpha}_j,k} + \mathsf{I}_{i,\mathsf{\beta}_j,k}\right) + \mathsf{\xi}_{i,j,k}\mathsf{I}_{i,4+j,k},
\end{align}
where
\begin{align} \label{equation: flux interference}
    \mathsf{\xi}_{i,j,k} = 6\mathsf{\gamma}_{i,j,k}\sqrt{\mathsf{F}_{i,\alpha_j,k}\mathsf{F}_{i,\beta_j,k}}.
\end{align}
The uncombined signals $\bm{\mathsf{I}}(t,\lambda) \in \mathbb{R}_{+}^{6\times10\times4}$ from the beam combiner output are
\begin{align} \label{equation: non-combined signals}
    \mathsf{I}_{i,l,k} = 
    \begin{cases}
        \left|\mathsf{E}_{i,l,k}\right|^2 & l = 1,...,4 \\
        \left|\mathsf{E}_{i,\mathsf{\alpha}_{l-4},k}\right|^2 + \left|\mathsf{E}_{i,\mathsf{\beta}_{l-4},k}\right|^2 +\\
        2 \Re\left\{\left(\mathsf{E}_{i,\mathsf{\alpha}_{l-4},k}\right)^{\dagger}\mathsf{E}_{i,\mathsf{\beta}_{l-4},k}e^{\mathrm{i}(\mathsf{\phi}_{\mathrm{eff},i}\frac{\lambda_\mathrm{eff}}{\lambda}+\Delta\mathsf{\Phi}_{i,l-4,k})}\right\} & l = 5,...,10,
    \end{cases}
\end{align}
with  
\begin{align} \label{equation: individual signal}
    \mathsf{E}_{i,j,k} = \sqrt{\mathsf{F}_{i,j,k}}\mathsf{\eta}_{j}f_{\mathrm{telluric}}.
\end{align}
Here  $\bm{\mathsf{\phi}}_{\mathrm{eff}}(t) \in (0,2\pi)^{6}$ are the recorded FT residual piston errors.

Equations (\ref{equation: output signals})-(\ref{equation: individual signal}) describe how individual telescope fluxes, fiber coupling, atmospheric transmission, and residual piston errors combine to produce the final interferometric signals used by \texttt{SIMTERFERE}.

\subsection{Hyperparameters} \label{subsection: hyper parameter}
The hyperparameters are defined by the user to set the input stellar spectrum, specify telluric properties, adjust temporal and spectral resolution, modify sub-instrument performances, apply noise to the signals, and configure the V2PMs.

For this work, we adopted the \texttt{NewEra PHOENIX} model grid \citep{hauschildt_newera_2025} with an effective temperature of 8000\,K, a logarithmic surface gravity of log\,g = 4\,cgs, and solar metallicity, approximately matching the properties of \textbeta\ Pictoris. We applied rotational broadening of 130\,km/s \citep{royer_rotational_2007} and a wavelength shift of -60\,km/s or -0.45\,nm to shift the Br\textgamma\ absorption from vacuum to air \citep[-80\,km/s or -0.6\,nm]{mathar_refractive_2007} and correct for the systematic velocity in the observed spectra \citep[20\,km/s or 0.15\,nm]{gontcharov_pulkovo_2006}. We further binned the spectrum to match the GRAVITY SC and FT spectral resolution.

For telluric absorption, we empirically adjusted molecular abundances to match the measured spectra: H\textsubscript{2}O at 130\% of the measured ambient relative humidity, used as a proxy for the total abundance; CO\textsubscript{2} at 125\%; and CH\textsubscript{4} at 140\% of the default \texttt{LBLRTM} values. We left all other molecular abundances at their default settings.

The simulation subsamples each integration with 100 frames and OPDs. We empirically find that increasing the number of samples beyond 100 does not significantly change the simulated coherent fluxes. The simulation computes wavefronts and OPDs at this high cadence, whereas other quantities (e.g., dispersion and transmission) are evaluated only once per exposure. 

\begin{figure*}[!t]
    \sidecaption
    \includegraphics[width=12cm]{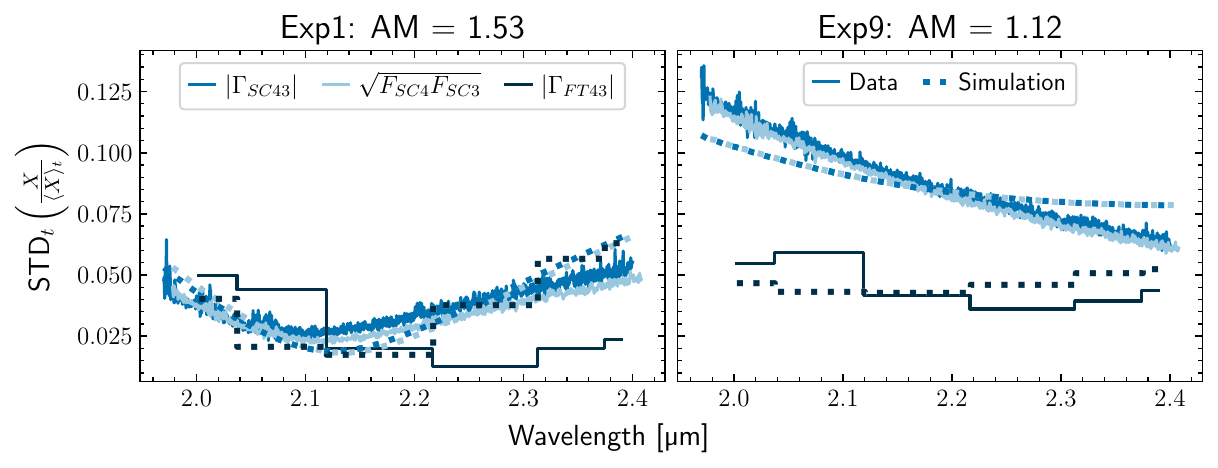}
    \caption{\textit{Left:} Standard Deviation (STD) of the measured (solid lines) and simulated (dotted lines) relative SC (1628 spectral bins) absolute coherent flux $|\Gamma_{\mathrm{SC43}}|$, the combined SC flux $\sqrt{F_{\mathrm{SC}4}F_{\mathrm{SC}3}}$, and the FT (6 spectral bins) absolute coherent flux $|\Gamma_{\mathrm{FT43}}|$ (different shades) for an examples exposure at high airmass on baseline 43. \textit{Right:} Same as the left panel, but for an exposure at low airmass.}
    \label{figure:normalized_st_am_43}
\end{figure*}

No modifications to sub-instrument performance or additional noise are applied, allowing the focus to remain on systematic effects.

The simulation initializes the SC V2PM using the measured and normalized recorded V2PM from daytime calibration, consistent with the GRAVITY pipeline. It is then rescaled using the measured throughput for each baseline and wavelength stored in the calibration file’s C-Matrix. The FT V2PM is modeled as an ideal beam combiner.

\section{Results} \label{section: results}
To characterize the transmissivity of GRAVITY and identify the sources of systematic effects, we simulated all on-star observations (18 exposures with each $16 \times 3\,s$ integration) using the configuration described in Sect. \ref{subsection: hyper parameter}. By comparing these controlled simulations to the measured data, we can disentangle the contributions of instrumental, atmospheric, and fiber-coupling effects that are otherwise inseparable in the raw observations.

Figure \ref{figure: visibility_example} shows the measured and simulated absolute coherent fluxes for an example integration at the beginning of the night. A movie comprising all integrations of this exposure is available online. The simulations qualitatively reproduce the absolute flux levels, which are dominated by the combination of instrumental and atmospheric transmission.

Relative absolute coherent fluxes exhibit low-order variations on the order of 10\%, reflecting systematic throughput fluctuations. These variations are primarily caused by time-dependent changes in fiber coupling, amplified by atmospheric dispersion. They persist across all exposures and are well modeled by second-order wavelength polynomials $Q_{\lambda,2}$. Consequently, we can express the total effective transmissivity as
\begin{align}
    G(\bm{\mathsf{b}},t,\lambda) \approx Q_{\lambda,2}(\bm{\mathsf{b}},t,\lambda) \cdot G_{\mathrm{instrument}}(\bm{\mathsf{b}},\lambda) \cdot G_{\mathrm{telluric}}(t,\lambda).
\end{align}
The increased scatter in the measured normalized coherent fluxes arises from photon noise and chromatic dispersion, particularly in strong telluric absorption bands. We did not include these additional noise terms in the simulations. 

\subsection{Throughput variations} \label{subsection: systematic effects}
To quantify the impact of systematic low-order throughput variations on GRAVITY observations, we first examined variations within individual exposures and subsequently across multiple exposures.

\subsubsection{Variations between integrations} \label{subsubsection: t1s}
Figure \ref{figure:normalized_st_am_43} shows the standard-deviations (STDs) of the measured and simulated normalized SC and FT absolute coherent fluxes, as well as combined SC flux-pairs, for two representative exposures at high and low airmass for baseline 43. We provide the complete plot with all baselines in Fig. \ref{figure:normalized_st_am}.

The simulations qualitatively reproduce the observed variations, capturing both the wavelength dependence and the airmass-related gradient introduced by throughput fluctuations. The FT generally shows better agreement with the simulations, whereas the SC exhibits higher STDs. We attribute this discrepancy primarily to fiber position offsets; we discuss this in more detail in Sect. \ref{subsection: fiber offsets}.

The STDs of the absolute coherent fluxes closely match those of the combined fluxes. This similar behavior indicates that a common source causes the variations across all four telescopes. We observe residual excess noise in strong absorption bands, which likely originates from uncorrected chromatic dispersion. We therefore conclude that systematic throughput variations in each telescope flux are the primary source of scatter in both absolute and combined coherent fluxes.

The wavelength dependence of the STDs also exhibits a clear airmass effect: at high airmass, STDs show pronounced “v-shaped” gradients centered around the effective fiber coupling wavelength, reflecting increased atmospheric dispersion. At low airmass, the STDs decrease slightly with wavelength because reduced wavefront errors produce more stable PSFs.

\subsubsection{Fiber offsets} \label{subsection: fiber offsets}

\begin{figure}[!h]
    \centering
    \includegraphics[width=0.9\hsize]{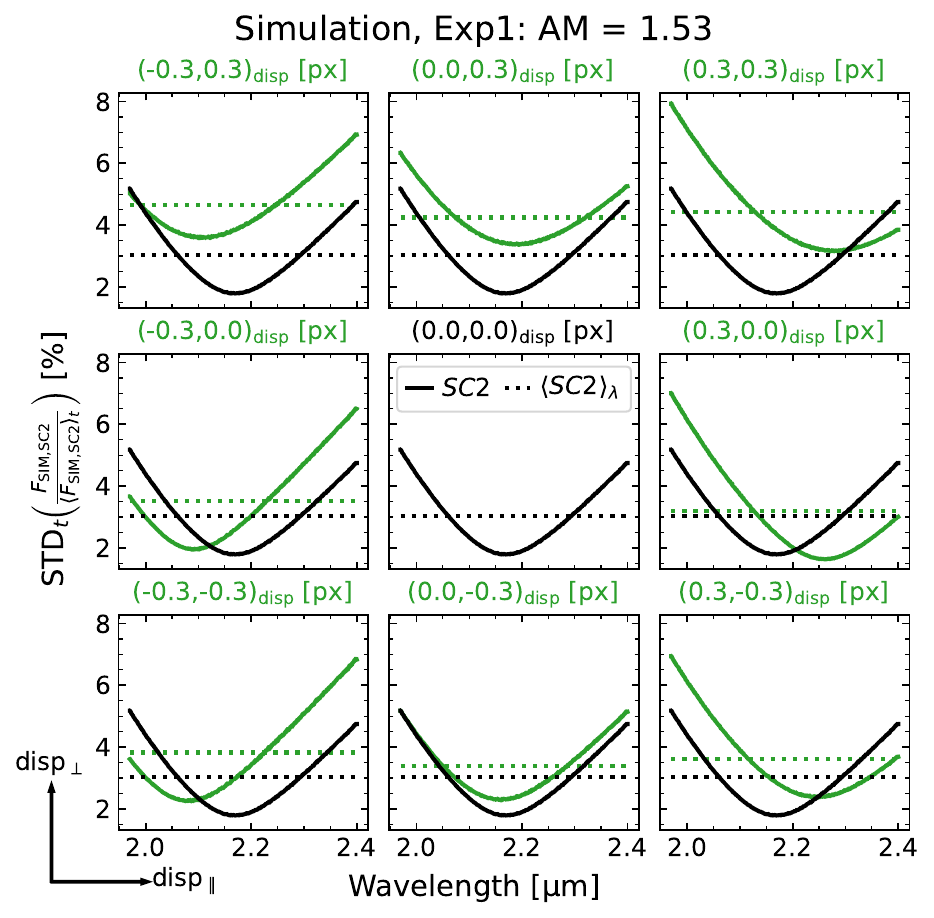}
    \caption{Standard deviation (STD) of the simulated SC2 relative fluxes for different applied fiber offsets, either along or orthogonal to the dispersion direction. Black lines represent STDs with manual fiber tracking and no applied offsets, while green lines show STDs with applied offsets. The dotted lines indicate the respective wavelength average. An offset parallel to the dispersion direction shifts the characteristic ``V-shape'' toward positive wavelengths, and an antiparallel shift produces the opposite effect. Offsets orthogonal to the dispersion direction increase the overall STDs. The observed asymmetries between opposite offset directions arise from the asymmetry of the PSFs.}
    \label{figure: fiber offsets}
\end{figure}

Figure \ref{figure: fiber offsets} illustrates that fiber positions play a critical role in the observed STD systematics. In the simulation, we computed fluxes using manually measured PSF centroids at 2.18\,μm for each integration, rather than using the recorded fiber positions.  We applied sub-pixel offsets (corresponding to $\approx 5\,\mathrm{mas}$) to further probe the sensitivity of throughput to fiber alignment.

\begin{figure*}[!t]
    \centering
    \includegraphics[width=0.75\textwidth]{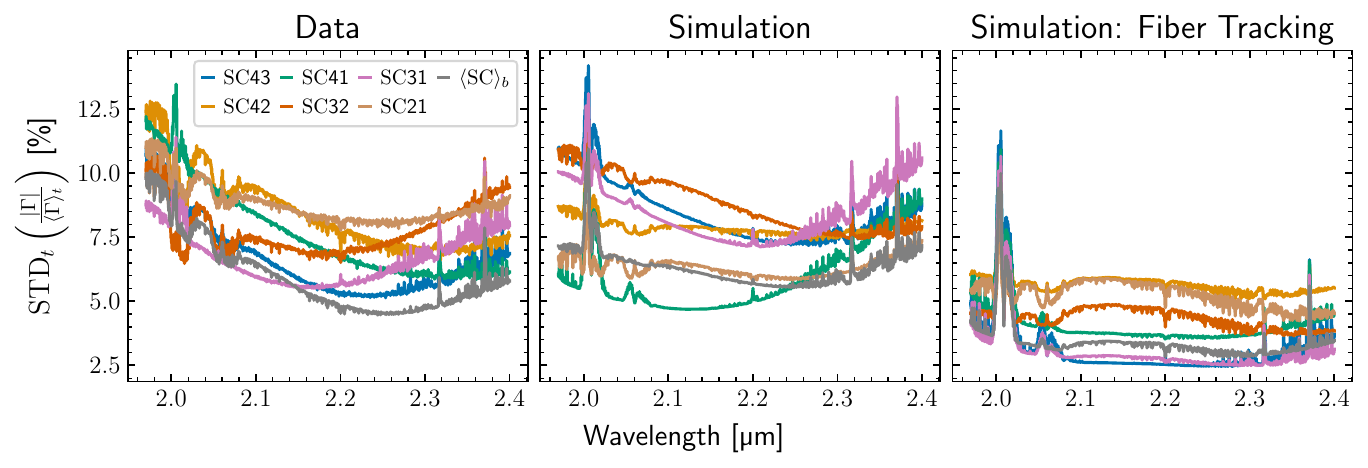}
    \caption{\textit{Left:} Standard deviation (STD) of the measured relative absolute coherent SC fluxes for all exposures during the observation, shown for the six baselines (43, ..., 21) and the baseline average $\langle \cdot \rangle_{b}$ (different colors). \textit{Middle:} Same as the left panel, but showing the respective simulation results. \textit{Right:} Same as the middle panel, but with manual fiber tracking applied during the simulations.}
    \label{figure: normalized std full}
\end{figure*}

Offsets along the atmospheric dispersion direction shift the characteristic v-shaped STD pattern along the wavelength axis, while offsets orthogonal to the refraction direction increase the overall STD amplitude. These effects explain much of the discrepancy between the SC, FT, and simulated STDs, indicating that even small fiber misalignments can dominate systematic throughput variations.

Quantitatively, the STD shape and amplitude are extremely sensitive to fiber position: an offset of just 0.3\,pixel on the acquisition camera can increase the variations by more than 50\%. These findings suggest that the recorded fiber positions do not fully reflect the actual positions during the observations. The origin of these offsets remains unclear and warrants further investigation.

Importantly, the mismatch between the SC and FT fiber positions prevents direct calibration of the SC fluxes using the FT.

\subsubsection{Variations between exposures} \label{subsubsection: t1h}
Following the approach in Sect. \ref{subsubsection: t1s}, we computed the STDs of the measured and simulated normalized absolute coherent fluxes across all exposures over the full night. We also evaluated simulations with manual fiber tracking  to assess the impact of fiber alignment.

Figure \ref{figure: normalized std full} shows that systematic throughput variations dominate on hour-long timescales, while telluric variations contribute only secondary, smaller fluctuations. The overall systematic variations of $\sim 10\%$ are consistent with previously reported multi-epoch differences in \texttt{exogravity} spectra \citep{kammerer_exogravity_2025}.

The measured and simulated STDs agree qualitatively, but individual baselines show differing offsets and trends. Comparison of simulations with and without manual fiber tracking demonstrates that these discrepancies primarily result from misaligned fiber positions. Under ideal fiber tracking, observed throughput variations could be reduced by more than a factor of two.

\subsection{Low wind effect} \label{subsection: lwe}
\begin{figure}[!h]
    \centering
        \centering
        \includegraphics[width=0.9\hsize]{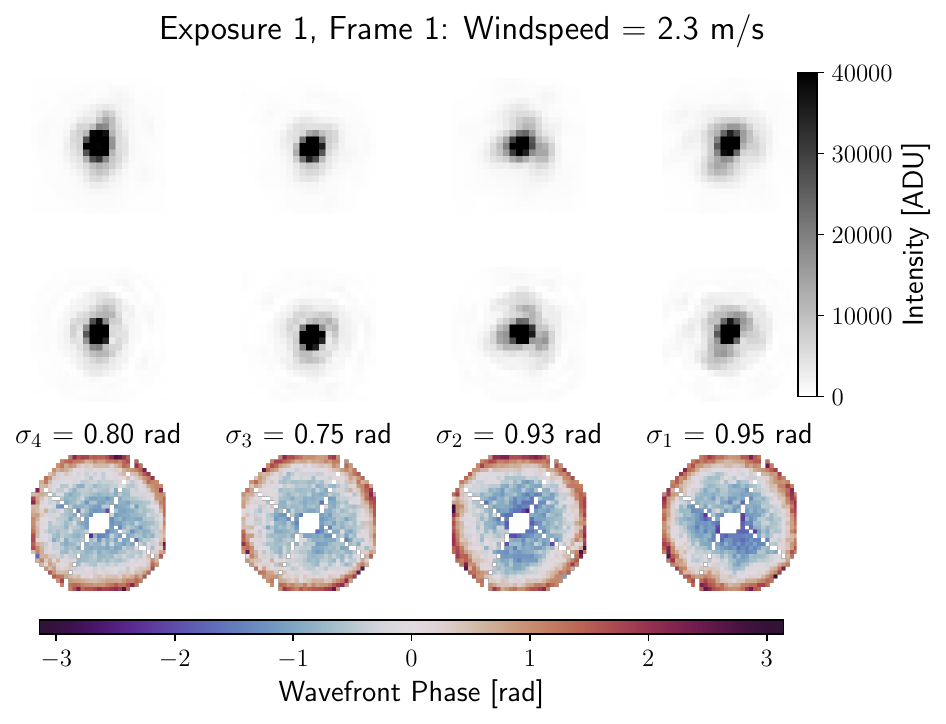}
        \caption{\textit{Top:} Example PSF frames of the four UTs (from left to right: 4, ..., 1) from the acquisition camera at moderate wind speed, saturated at 40000\,ADU. \textit{Center:} PSF fits to the data. \textit{Bottom:} Retrieved tip-tilt-subtracted wavefronts from the PSF fits. The wavefront errors are measured using the standard deviation. The associated movie is available online.}
        \label{figure: wavefront good}
\end{figure}

\begin{figure}[!h]   
        \centering
        \includegraphics[width=0.9\hsize]{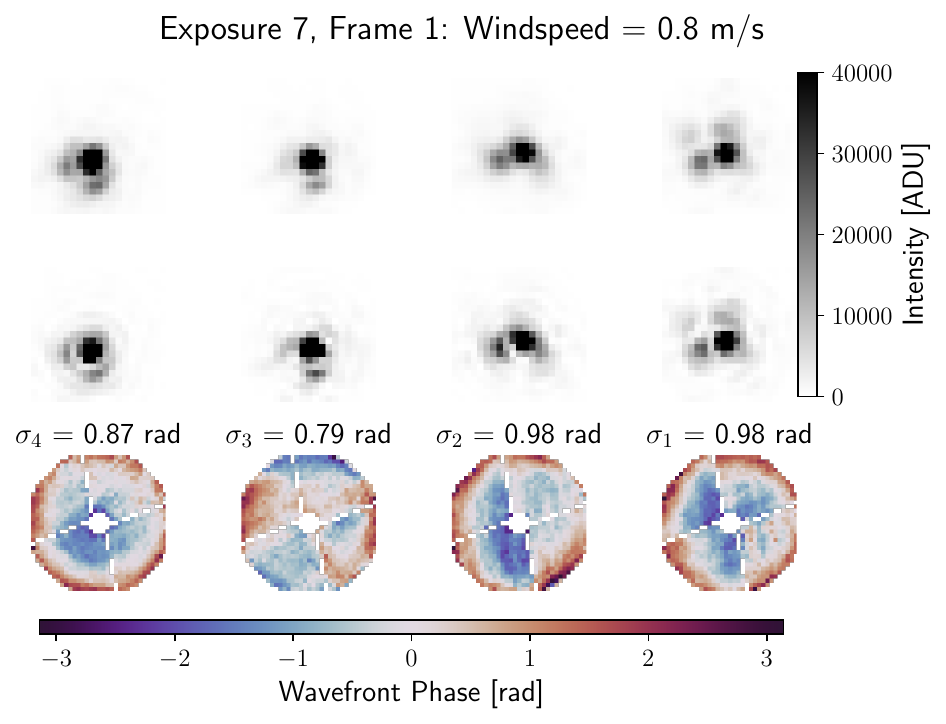}
        \caption{Same as Fig. \ref{figure: wavefront good}, but at very low wind speed. Strong phase discontinuities between the spiders are present in the UT1 and UT2 wavefronts, introducing pronounced side lobes in the PSFs. The associated movie is available online.}
        \label{figure: wavefront bad}
\end{figure}

Throughout the night, low wind speeds ($< 3\,\mathrm{m/s}$) degraded the quality and stability of the PSF due to the LWE. Figures \ref{figure: wavefront good} and \ref{figure: wavefront bad} show example PSF and wavefront fit frames under moderate and strong LWE conditions. Movies comprising all frames for the two exposures are available online. We derived the PSF fits using focal-plane wavefront sensing within \texttt{SIMTERFERE}. As the LWE strengthens, phase discontinuities appear between the four spider quadrants, causing increased light leakage into the side lobes of the PSF. This effect is particularly pronounced for UT1 and UT2, which lack specialized spider coatings \citep{pourre_low-wind-effect_2022}.


\begin{figure}[h]
    \centering
    \includegraphics[width=0.9\hsize]{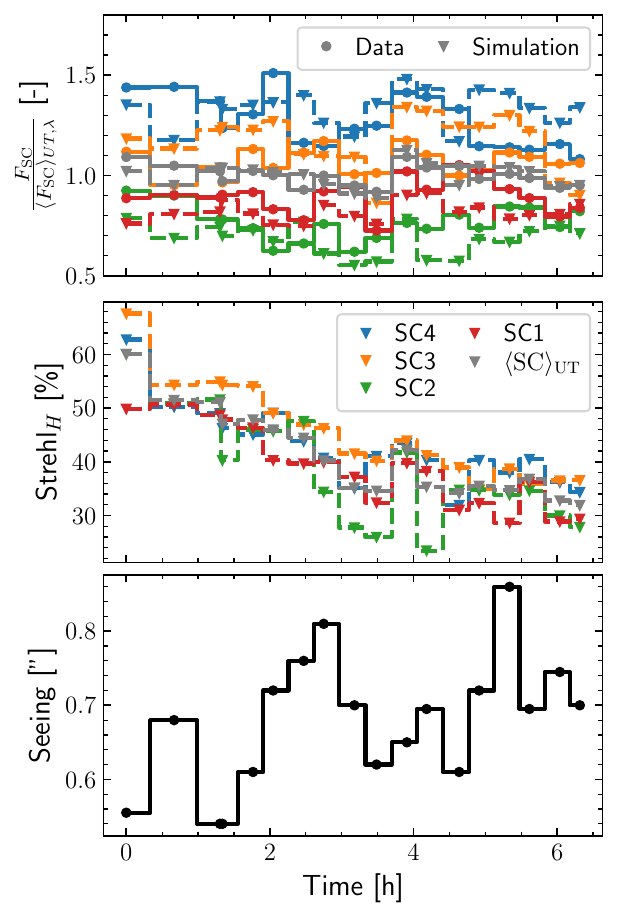}
    \caption{\textit{Top:} Measured (dots, solid lines) and simulated (triangles, dashed lines) relative SC fluxes of the four UTs (4, ..., 1) and their average (different colors) for all exposures during the observation. \textit{Center:} Respective H-band Strehl ratios derived from focal-plane wavefront sensing. \textit{Bottom:} Recorded seeing during the observing night.}
    \label{figure: wavefront flux}
\end{figure}

Figure \ref{figure: wavefront flux} presents the measured and simulated relative UT fluxes for each exposure. The simulations reproduce the overall trends and agree reasonably well with the individual UTs and the average fluxes. The remaining differences are primarily due to misaligned fiber positions, as discussed in Sect. \ref{subsection: fiber offsets}.

The flux trends show an overall complex behavior influenced by airmass, AO performance, and fiber positioning. The derived Strehl ratio, representative of AO performance, only loosely correlates with the seeing conditions because the LWE further degrades it. The UT3 and UT4 telescopes consistently exhibit higher Strehl ratios than UT1 and UT2, reflecting the improved PSF stability provided by the spider coatings.  
 
\subsection{Polynomial correction} \label{subsection: continuum correction}
As described in Sect. \ref{section: introduction}, we extracted exoplanet spectra using the stellar spectrum measured immediately before and after each observation in \texttt{exogravity}. To quantify the precision of this approach, we computed the root-mean-square (RMS) relative differences of the absolute coherent fluxes between consecutive star exposures.

As shown in Sect. \ref{section: results}, the relative throughput variations can be modeled as low-order polynomials in wavelength. We therefore applied a throughput correction by fitting second-order polynomials $Q_{\lambda,2}(\bm{b},t,\lambda) \in \mathbb{R}$:
\begin{align}
    Q_{\lambda,2} \approx \frac{|\Gamma|}{\langle|\Gamma|\rangle_{t}},
\end{align}
which we then applied to the absolute fluxes to obtain the throughput corrected coherent fluxes $|\Gamma|_Q(\bm{b},t,\lambda) \in \mathbb{C}$:
\begin{align}
    |\Gamma|_Q = \frac{|\Gamma|}{Q_{\lambda,2}}.
\end{align}

\begin{figure*}[t]
    \sidecaption
    \includegraphics[width=12cm]{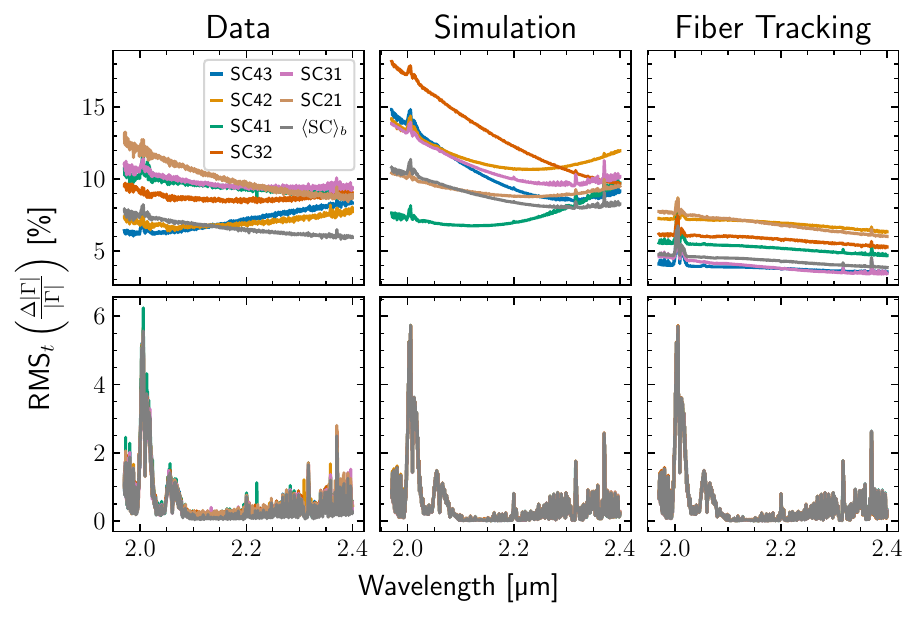}
    \caption{\textit{Top:} Root-mean-square (RMS) of the relative absolute SC coherent flux between two subsequent exposures, shown per wavelength bin for the data (left), the simulations (center), and the simulations with applied fiber tracking, for the six baselines (43, ..., 21) and the baseline average $\langle \cdot \rangle_{b}$ (different colors). \textit{Bottom:} Same as top panel, but after applying the polynomial corrections.}
    \label{figure: rms visibility}
\end{figure*}

Figure \ref{figure: rms visibility} shows the measured and simulated relative SC RMS coherent flux differences before and after polynomial calibration. The mean relative RMS decreases from $6.5^{+0.7}_{-0.4}\%$ to $0.21^{+0.49}_{-0.13}\%$, with the residuals dominated by telluric variations.

Although the simulated fluxes with and without manual fiber tracking exhibit markedly different RMS trends prior to correction, the residuals converge after polynomial calibration. This result demonstrates the effectiveness of low-order polynomial correction in mitigating systematic throughput variations. 

\subsection{Telluric variations} \label{subsection: continuum correction}
To quantify residual variations after polynomial correction, we measured the baseline- and wavelength-averaged relative change in absolute coherent flux between consecutive polynomial-corrected exposures as a function of airmass change. We compared the measured and simulated data with a pure telluric model computed with the same atmospheric parameters used in the simulations.

\begin{figure}
    \centering
    \includegraphics[width=0.9\hsize]{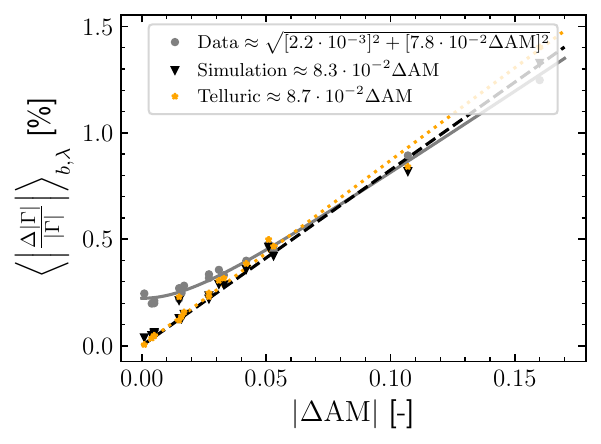}
    \caption{Baseline- and wavelength-averaged relative absolute coherent flux change between two subsequent stellar pointings, shown as a function of absolute airmass change. The simulations (triangles, dashed lines) and the pure telluric models (stars, dotted lines) do not include noise, which results in a linear behavior. For the data (dots, solid lines), the flux change instead converges toward the noise floor at small airmass changes.}
    \label{figure: relative change airmass}
\end{figure}

Figure \ref{figure: relative change airmass} shows that the coherent flux increases approximately linearly for large airmass changes but converges to the noise floor for small airmass changes. As expected, this noise floor is absent in both the simulations and the pure telluric model. We fit the behavior with a combination of a constant noise term and a linear term proportional to airmass change:
\begin{align}
    \left|\frac{\Delta|\Gamma|}{|\Gamma|}\right|(\lambda,\Delta\mathrm{AM}) = \sqrt{\mathrm{SNR}^{-2}_{|\Gamma|}(\lambda)+\left[\frac{\partial|\Gamma|}{\partial\mathrm{AM}\cdot |\Gamma|}(\lambda) \cdot \Delta\mathrm{AM}\right]^{2}}.
\end{align}

The three curves show good agreement with the fitted curves, with scattering dominated by non-airmass-dependent variations such as changes in pressure, temperature, and water vapor. Deviations at the largest airmass changes are attributed to small biases introduced by the polynomial correction.

\begin{figure}
    \centering
    \includegraphics[width=0.9\hsize]{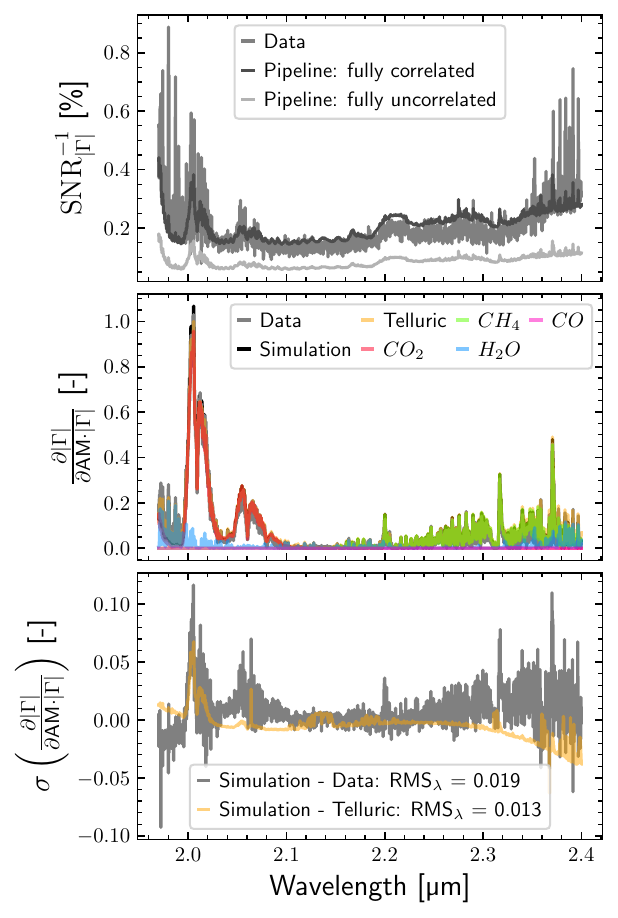}
    \caption{\textit{Top:} Relative absolute coherent flux noise as a function of wavelength, measured in the data and estimated by the GRAVITY pipeline assuming fully correlated and uncorrelated baselines. \textit{Center:} Relative absolute coherent flux slopes as a function of wavelength for the data, the simulations, and the pure telluric models, including their respective molecular contributions. \textit{Bottom:} Residuals of the relative absolute coherent flux slopes between the simulations, the data, and the telluric models.}
    \label{figure: noise wavelength}
\end{figure}

Following Beer–Lambert’s law, the slope with respect to airmass directly measures the telluric transmissivity at unit airmass:
\begin{align}
    \frac{\partial|\Gamma|}{\partial\mathrm{AM}\cdot |\Gamma|}(\lambda) = \frac{\partial\ln{G_{\mathrm{telluric},\mathrm{AM}=1}}}{\partial AM}(\lambda) = \ln{G_{\mathrm{telluric},\mathrm{AM}=1}(\lambda).}
\end{align}
Figure \ref{figure: noise wavelength} shows the wavelength-dependent measured relative noise $\mathrm{S/N}^{-1}$ and airmass slopes $\partial|\Gamma|/\partial\mathrm{AM}\cdot |\Gamma|$. The measured noise clearly lies above the pipeline prediction assuming no correlation (i.e., $1/\sqrt{6}$ scaling) and is only slightly below the fully correlated noise estimate at intermediate wavelengths. The increase at the low and high ends results primarily from polynomial correction biases.

These results highlight the strong correlation among the six baseline coherent fluxes, which must be taken into account in noise estimation and propagation for GRAVITY data.

Simulations, measurements, and pure telluric models show excellent agreement, indicating that polynomial-corrected fluxes are dominated by telluric variations. The measured RMS residuals of 0.019 and 0.013 correspond to relative flux errors of 0.05\% and 0.035\% at median airmass changes, well below the intrinsic data noise.

In the K band, the main telluric absorbers are CO\textsubscript{2}, CH\textsubscript{4}, and H\textsubscript{2}O: CO\textsubscript{2} absorbs mainly below 2.1\,µm, CH\textsubscript{4} above 2.2\,µm, and H\textsubscript{2}O below 2\,µm and above 2.3\,µm. We included CO, an important exoplanet atmosphere tracer, as a reference. The CO telluric absorption from 2.3\,µm to 2.4\,µm (peak slope of 0.0035) is negligible when dealing with GRAVITY spectra. 

The residual discrepancies between the simulated and telluric slopes are caused by the imperfect polynomial correction. Variable telluric line depths, uncorrelated with the systematic throughput, bias the polynomial fits, particularly at the low- and high-wavelength ends. To minimize this effect, we fit the polynomials only at wavelengths where the modeled transmission exceeds 99\%, leaving the fits still imperfect.

\subsection{Implications for exoplanet spectroscopy}
As outlined in Sect. \ref{section: introduction}, the precision of the predicted on-star coherent flux during on-planet observations scales linearly with the retrieved contrast spectrum precision in \texttt{exogravity}. Quantifying both the accuracy and long-term stability of this prediction is therefore essential for high-fidelity exoplanet spectroscopy.

To evaluate prediction precision, we compared the measured polynomial-corrected absolute coherent fluxes $|\Gamma_{i}|$ with the predicted fluxes derived from the preceding and subsequent exposures $|\Gamma_{i-1/i+1}|$.

Following Eq. (\ref{equation: on-star prediction}), in the current implementation of \texttt{exogravity}, we obtain the predicted coherent flux  by averaging the absolute coherent fluxes before and after the science exposure:
\begin{align}
    |\Gamma_{i}| \approx |\Gamma_{\mathrm{Averaged}}| = \frac{|\Gamma_{i-1}|+|\Gamma_{i+1}|}{2}.
\end{align}
As derived in Sect. \ref{subsection: continuum correction}, we achieved a more accurate prediction by applying a weighted linear interpolation as a function of airmass:
\begin{align}
    |\Gamma_{i}| &\approx |\Gamma_{\mathrm{Interpolated}}| \nonumber \\
    &= w \cdot \left(\frac{(|\Gamma_{i+1}|-|\Gamma_{i-1}|)(\mathrm{AM}_i-\mathrm{AM}_{i+1})}{\mathrm{AM}_{i+1} - \mathrm{AM}_{i-1}} + |\Gamma_{i+1}|\right) + \nonumber\\
    & (1-w) \cdot \Gamma_{\mathrm{Averaged}},
\end{align}
where the weight factor $w \in (0,1)$ is defined as
\begin{align}
    w = \frac{(\mathrm{AM}_{i+1} - \mathrm{AM}_{i-1})^2}{\delta^2+(\mathrm{AM}_{i+1} - \mathrm{AM}_{i-1})^2}.
\end{align}
This weighting scheme suppresses numerical instabilities for very small airmass changes ($|\mathrm{AM}_{i+1} - \mathrm{AM}_{i-1}| \rightarrow 0$), while smoothly transitioning to interpolation for large changes. We empirically find $\delta \approx 0.03$ to provide good improvements for large airmass changes while still suppressing noise amplification for low changes. For airmass changes below 0.03, photon noise dominates over telluric variability, as illustrated in Fig. \ref{figure: relative change airmass}.

\begin{figure}[h]
    \centering
    \includegraphics[width=0.9\hsize]{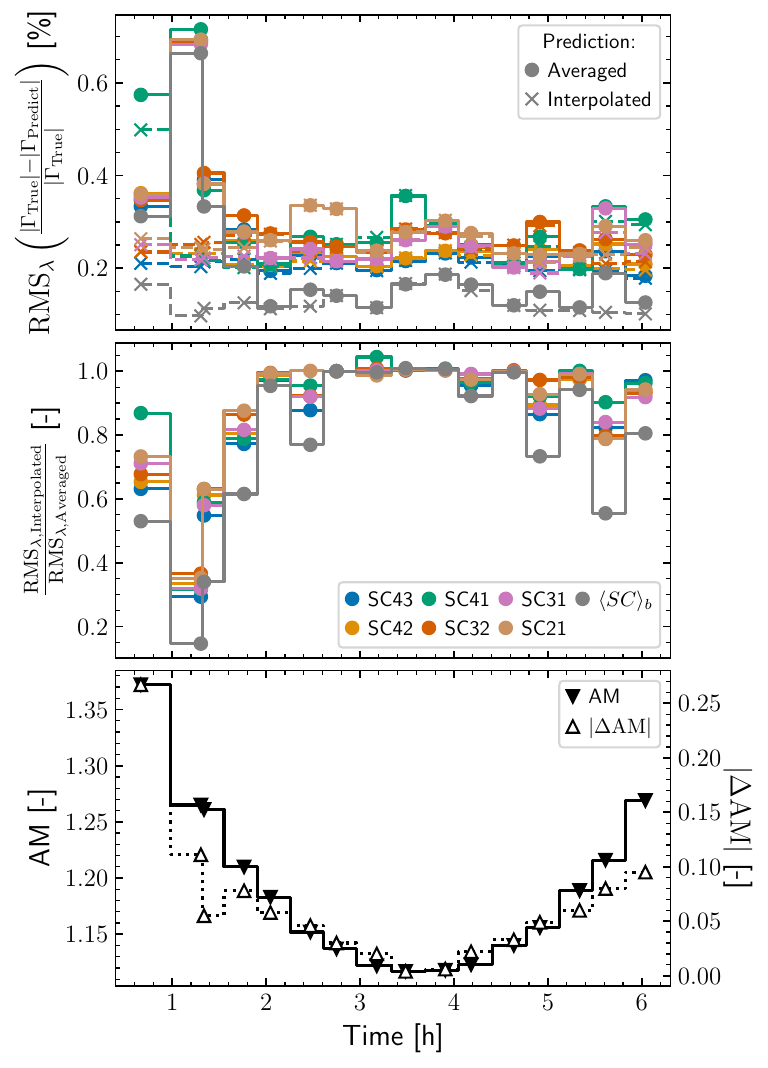}
    \caption{\textit{Top:} Wavelength RMS of the relative difference between the measured $\mathrm{True}$ and $\mathrm{Predicted}$ absolute coherent flux as a function of time. The RMS values are shown for each interferometric baseline and for the baseline average (different colors), and are computed for each on-star exposure with at least one preceding and subsequent calibration exposure, using averaging (dots, solid lines) and interpolation (crosses, dashed lines) approaches. \textit{Center:} Ratio between the RMS obtained with the averaging and interpolation methods for each exposure. \textit{Bottom:} Airmass (filled triangles, solid lines) at the time of observation and absolute airmass difference (open triangles, dotted lines) between the preceding and subsequent calibration exposures. Because the observations are obtained at approximately constant time intervals, the airmass differences are strongly correlated with the airmass.}
    \label{figure: stellar spectrum precision}
\end{figure}

We define the precision as the wavelength RMS of the relative difference between the measured $\mathrm{True}$ and $\mathrm{Predicted}$ absolute coherent flux. Figure \ref{figure: stellar spectrum precision} shows that the averaging approach exhibits a strong decrease in precision at high airmass or during periods of rapid airmass change, leading to increased temporal scatter. In contrast, the interpolation approach largely removes this dependence, yielding photon-noise-dominated residuals that remain stable throughout the full seven-hour observation for both individual baselines and the baseline-averaged fluxes. We observe no systematic drift over the full time series, demonstrating the long-term spectrometric stability of GRAVITY+ on multi-hour timescales.

\begin{table}[h]
\caption{Achieved absolute coherent flux precisions and corresponding ratios for the average and interpolation prediction method.}      
\label{table: spectrum precision}      
\centering               
\begin{tabular}{c|ccc}  
\hline\hline 
Baseline & Averaged [\%] & Interpolated [\%] & Ratio [-]\\
\hline 
SC43 & $0.224^{+0.089}_{-0.026}$ & $0.204^{+0.011}_{-0.009}$ & 0.91 \\
SC42 & $0.235^{+0.086}_{-0.029}$ & $0.215^{+0.017}_{-0.014}$ & 0.92 \\
SC41 & $0.27^{+0.10}_{-0.04}$ & $0.25^{+0.05}_{-0.04}$ & 0.93 \\
SC32 & $0.269^{+0.065}_{-0.028}$ & $0.248^{+0.027}_{-0.014}$ & 0.92 \\
SC31 & $0.25^{+0.09}_{-0.03}$ & $0.224^{+0.033}_{-0.008}$ & 0.90 \\
SC21 & $0.28^{+0.07}_{-0.04}$ & $0.245^{+0.059}_{-0.015}$ & 0.88 \\
$\langle \mathrm{SC} \rangle_b$ & $0.16^{+0.11}_{-0.04}$ & $0.117^{+0.044}_{-0.010}$ & 0.73 \\
\hline    
\end{tabular}
\tablefoot{The precision is defined as RMS temporal median with 16-84th percentile range uncertainties per baseline and baseline average. The ratio is given as the quotient between the median averaged and interpolated value.}
\end{table}

Table \ref{table: spectrum precision} summarizes the median prediction precision achieved across the full time series, together with the corresponding 16–84th percentile ranges. For individual baselines, the interpolation approach yields a consistent improvement of approximately 10\% compared to simple averaging. For the baseline-averaged fluxes, where the photon noise is further reduced, the improvement reaches approximately 30\%. In this case, the averaging approach achieves a median precision of $0.16^{+0.11}_{-0.04}\%$, while the interpolation approach reaches a precision of $0.117^{+0.044}_{-0.010}\%$.

This combination of high precision, low intrinsic scatter, and temporal stability over the continuous seven-hour observation demonstrates that the polynomial-corrected and airmass-interpolated coherent-flux prediction provides a robust foundation for high-fidelity, long-duration spectroscopic characterization of exoplanets with GRAVITY+.

\section{Limitations} \label{section: limitations}
As demonstrated in Sect. \ref{section: results}, \texttt{SIMTERFERE} qualitatively reproduces the absolute coherent fluxes and their short-term variations for our high-S/N ($\approx 180$), short-exposure (48\,s) \textbeta\ Pictoris observations. On these timescales, the approximations made for time-discretized wavefronts, atmospheric dispersion, and transmission are thus adequate.

However, several limitations remain.
\begin{enumerate}
    \item For longer exposures, atmospheric conditions (e.g., temperature, pressure, and water vapor) may vary during the integration, requiring a more detailed temporal treatment of wavefronts and telluric effects.
    
    \item \texttt{SIMTERFERE} does not compute detector images containing the 24 spectral traces, but instead computes the signals directly. As a result, optical effects (e.g., line spread functions or curved spectral traces) and detector effects (e.g., bad pixels or flat fields) that are corrected by the GRAVITY pipeline are not included. Nevertheless, the agreement between the data and the simulations verifies the excellent performance of the pipeline.
    
    \item  \texttt{SIMTERFERE} currently does not implement measurement noise (i.e. photon noise, background noise, and detector readout noise) or its propagation through the P2VM method. These effects become particularly significant for low-S/N targets, such as faint exoplanets. In Sect.~\ref{subsubsection: t1s}, we show that the dominant noise in the \textbeta\ Pictoris observations arises from chromatic dispersion, particularly in the CO\textsubscript{2} atmospheric absorption bands, which are also not included in \texttt{SIMTERFERE}.
    
    \item The current implementation supports only on-star observations in the dual-field on-axis mode. Supporting on-planet simulations would require the implementation of a two-source mode, in which the stellar and planetary contributions are simulated separately and then combined. Additional optical effects such as anisoplanatism or anisopistonism would also need to be taken into account.
\end{enumerate}

\section{Discussion and conclusion} \label{section: discussion and conclusion}
The on-star coherent fluxes play a central role in measuring exoplanet contrast spectra. First, they provide the linear reference against which the on-planet coherent fluxes are fitted. Second, any systematic effects that influence the stellar fluxes are expected to influence the planetary fluxes. A robust characterization of the stellar coherent fluxes is therefore essential for quantifying the fidelity of extracted exoplanet spectra.

Our continuous seven-hour GRAVITY observation of the \textbeta\ Pictoris system offers a unique dataset for this purpose: a high-S/N, non-variable target that enables a systematic decomposition of the temporal and chromatic effects in the coherent fluxes. To interpret these data, we developed the data-driven simulation framework \texttt{SIMTERFERE}, which reproduces the observed flux behavior using the ancillary data recorded during the observations. This approach allows us to identify the physical origin of each systematic effect in isolation, which is otherwise inaccessible in the real data.  

Our analysis demonstrates that the dominant source of coherent-flux variations arises from non-constant single-mode fiber coupling. The coupling efficiency is set by the interplay between PSF stability, atmospheric dispersion, and fiber-position offsets. These effects produce wavelength-dependent throughput variations on the order of 10\%, a level consistent with the multi-epoch differences previously reported in \texttt{exogravity} exoplanet spectra \citep{kammerer_exogravity_2025}. Importantly, \texttt{SIMTERFERE} shows that these variations are systematic, not stochastic, and largely predictable from the measured AO performance.

We find that these coupling-induced variations manifest primarily as low-order chromatic changes, which can be effectively removed through second-order polynomial throughput corrections. Once this correction is applied, the remaining inter-exposure variability is dominated by atmospheric transmission changes. These telluric variations follow the Beer-Lambert law with high precision. Consequently, instead of the current \texttt{exogravity} pipeline approach of averaging the on-star exposures taken before and after the planet observation, a linear interpolation with airmass provides a more accurate and physically consistent reference for the stellar coherent flux.

After polynomial correction, the residual variations in the coherent fluxes fall to the level expected from telluric changes and noise, with no evidence for significant additional systematics. This demonstrates the excellent performance of the GRAVITY instrument and pipeline and shows that coherent-flux stability is fundamentally limited by atmospheric transmission rather than instrumental drifts once fiber-coupling systematics are removed.

We demonstrate that the GRAVITY+ instrument achieves exceptional long-term on-star stability, allowing the stellar coherent flux to be predicted with a precision of approximately 0.1\% over the entire seven-hour observing sequence. This precision is limited only by photon noise, highlighting the capability of the instrument to support high-fidelity exoplanet spectroscopy over multi-hour observations.

In forthcoming work, we will extend our analysis to the on-planet observations within our \textbeta\ Pictoris dataset. By extending and applying \texttt{SIMTERFERE}, we will test the \texttt{exogravity} pipeline end-to-end, quantify possible biases in the planet extraction, and evaluate the impact of residual stellar and telluric systematics on the recovered planet spectrum.

\section*{Data availability}
Supplementary data underlying the results presented in this article are available at \url{https://zenodo.org/records/17817680}. The \texttt{SIMTERFERE} framework is publicly available at \url{https://github.com/Manaffel/SIMTERFERE}.

\begin{acknowledgements}
The authors thank the anonymous referee for the constructive comments and suggestions, which helped to improve the quality of this work. This work made use of the Python packages: Astropy \citep{astropy_collaboration_astropy_2013,astropy_collaboration_astropy_2018,astropy_collaboration_astropy_2022}, Numpy \citep{harris_array_2020}, LBLRTM/TelFit \citep{clough_line-by-line_1992,clough_atmospheric_2005,gullikson_correcting_2014}, Matplotlib \citep{hunter_matplotlib_2007}, Poppy \citep{perrin_poppy_2016}, PyTorch \citep{paszke_pytorch_2019}, Scipy \citep{virtanen_scipy_2020}, and Seaborn \citep{waskom_seaborn_2021}. Based on observations collected at the European Southern Observatory under ESO programmes 114.27JS.
\end{acknowledgements}

\bibliographystyle{aa} 
\bibliography{SIMTERFERE} 

\begin{appendix}
\onecolumn
\section{Baseline-dependent systematic throughput variations}
    \begin{figure*}[h]
        \centering
        \includegraphics[width=0.75\textwidth]{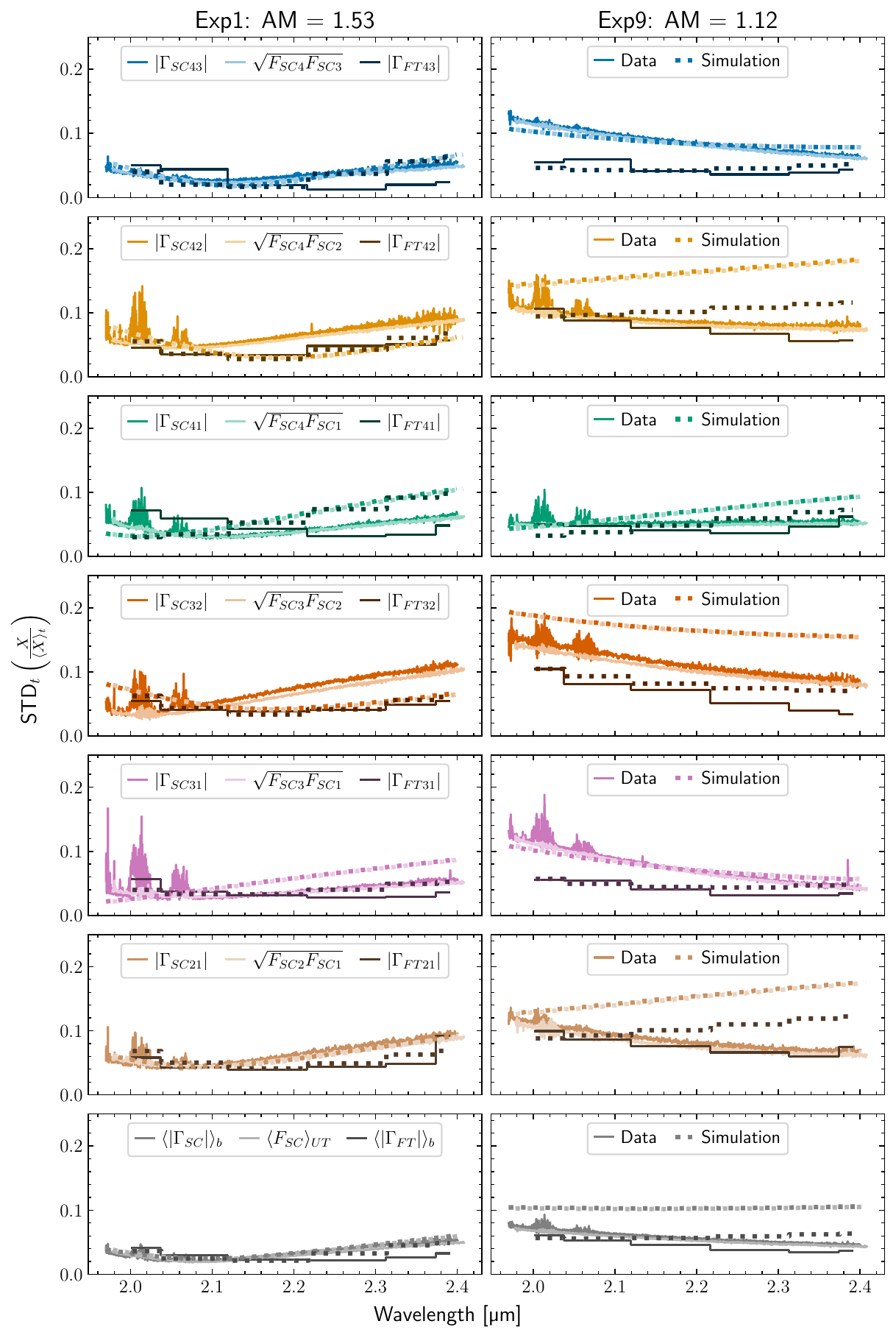}
        \caption{\textit{Left:} Standard deviation (STD) of the measured (solid lines) and simulated (dotted lines) relative SC (1628 spectral bins) absolute coherent fluxes $|\Gamma_{\mathrm{SC43}}|$, ..., $|\Gamma_{\mathrm{SC21}}|$, combined SC fluxes $\sqrt{F_{\mathrm{SC}4}F_{\mathrm{SC}3}}$, ..., $\sqrt{F_{\mathrm{SC}2}F_{\mathrm{SC}1}}$, and FT (6 spectral bins) absolute coherent fluxes $|\Gamma_{\mathrm{FT43}}|$, ..., $|\Gamma_{\mathrm{FT21}}|$ (different shades) for an example exposure at high airmass, shown for the six baselines (43, ..., 21) and their respective baseline averages $\langle \cdot \rangle_{b}$ (different colors from top to bottom). \textit{Right:} Same as the left panels but for an exposure at low airmass.}
        \label{figure:normalized_st_am}
    \end{figure*}
\end{appendix}

\end{document}